\documentclass[12pt,twoside,fleqn]{article}
\usepackage{espcrc1,epsfig,rotating}

\newcommand{\AmS}{{\protect\the\textfont2
  A\kern-.1667em\lower.5ex\hbox{M}\kern-.125emS}}

\begin{document}
\title{Parity Violation in Proton-Proton Scattering}

\author{
A.R.~Berdoz$^{a,b}$, 
J.~Birchall$^a$, 
J.D.~Bowman$^c$,
J.R.~Campbell$^a$, 
C.A.~Davis$^d$, 
A.A.~Green$^a$,
P.W.~Green$^e$,
A.A.~Hamian$^a$,
D.C.~Healey$^d$, 
R.~Helmer$^d$,
S.~Kadantsev$^f$,
Y.~Kuznetsov$^f$,  
R.~Laxdal$^d$,
L.~Lee$^a$, 
C.D.P.~Levy$^d$,
R.E.~Mischke$^c$,
S.A.~Page$^a$, 
W.D.~Ramsay$^a$, 
S.D.~Reitzner$^a$, 
G.~Roy$^e$, 
P.W.~Schmor$^d$,
A.M.~Sekulovich$^a$, 
J.~Soukup$^e$, 
G.M.~Stinson$^e$,
T.~Stocki$^e$,
V.~Sum$^a$,
N.A.~Titov$^f$, 
W.T.H. van Oers$^a$,
R-J.~Woo$^a$, 
A.N.~Zelenski$^f$\\
\vspace{0.4cm}%
$^a$Department of Physics, University of Manitoba, Winnipeg MB,
    Canada R3T 2N2\\
$^b$Department of Physics, Carnegie Mellon University, Pittsburgh PA, USA 15213\\
$^c$Los Alamos National Laboratory, Los Alamos NM, USA 87545\\
$^d$TRIUMF, 4004 Wesbrook Mall, Vancouver, B.C., Canada V6T 2A3\\
$^e$Department of Physics, University of Alberta, Edmonton AB, Canada T5G 2N5\\
$^f$Institute for Nuclear Research, Russian Academy of Sciences, 117312 Moscow,
Russia\\ 
\begin{center}
Presented by Willem T.H. van Oers
\end{center}}
 
\maketitle

\begin{abstract}
\begin{center}
Abstract
\end{center}

   Measurements of parity-violating longitudinal analyzing powers (normalized
asymmetries) in polarized proton-proton scattering provide a unique window on
the interplay between the weak and strong interactions between and within
hadrons.  Several new proton-proton parity
violation experiments are presently either being performed or are being
prepared for execution in the near future: at TRIUMF at 221 MeV and 450 MeV
and at COSY (Kernforschungsanlage J\"ulich) at 230 MeV and near 1.3 GeV. These
experiments are intended to provide stringent constraints on the set of six
effective weak meson-nucleon coupling constants, which characterize the weak
interaction between hadrons in the energy domain where meson exchange models
provide an appropriate description. The 221 MeV is unique in that it selects
a single transition amplitude ($^3$P$_2$ - $^1$D$_2$) and consequently constrains the
weak meson-nucleon coupling constant $h_\rho^{pp}$ .
  The TRIUMF 221 MeV proton-proton parity violation experiment is
described in some detail. A preliminary result for the longitudinal analyzing
power is $A_z$ = (1.1 $\pm$ 0.4 $\pm$ 0.4) $\times$ 10$^{-7}$. 
Further proton-proton parity violation experiments are commented on. The anomaly at 6 GeV/c
requires that a new multi-GeV proton-proton parity violation experiment
be performed.
\end{abstract}

\begin{center}
\rule{8.0cm}{0.2mm}
\end{center}
 

   One of the more promising ways to study the neutral weak current interaction
in hadronic systems is through measurements of parity violation in nucleon-nucleon 
(N-N) scattering. In the low-energy region, where meson exchange models
provide an adequate description of the strong N-N interaction, an extension can
be made to include the weak interaction. Pictorially, the exchanged meson
($\pi$, $\rho$, $\omega$) is emitted at a weak interaction vertex and absorbed at a
strong interaction vertex, or vice versa. The weak interaction vertex is
calculated from the Weinberg-Salam model with W- and Z-bosons exchanged between
intermediate mesons and constituent quarks, treating strong interaction effects
in renormalization group theory in the regime of nonperturbative QCD. In a
seminal paper following the above approach, restricted to one-boson exchanges,
Desplanques, Donoghue and Holstein (DDH) [1] have calculated a set of six weak
meson-nucleon coupling constants (a seventh was found to be rather small and is
usually neglected). These six weak meson-nucleon coupling constants are denoted:
 $f^1_\pi$, $h^0_\rho$, $h^1_\rho$, $h^2_\rho$, $h^0_\omega$, $h^1_\omega$; 
 where the subscript 
indicates the exchanged meson and the superscript the isospin change. DDH
tabulated ``best guess values" and ``reasonable ranges". As shown in Table 1,
the ``reasonable ranges of values" indicate uncertainties with respect to the
``best guess values" of a few hundred percent. Similar calculations have been
made by Dubovik and Zenkin (DZ) [2]. Extending the earlier work in the nucleon
sector, Feldman, Crawford, Dubach, and Holstein (FCDH) [3] included the weak
$\Delta$-nucleon-meson and weak $\Delta$--$\Delta$--meson parity violating vertices for
$\pi$, $\rho$, and $\omega$ mesons. The latter authors also present both ``best guess
values" and ``reasonable ranges" for the weak meson-nucleon coupling
constants. Using the expressions of an earlier paper by Desplanques (D) [4]
the latter authors (FCDH) present a third set of weak meson-nucleon coupling
constants. It is to be noted that the large ranges of possible theoretical
values persist. Taking into account the more recent experiments, Desplanques
[5] has argued for a reduced range for the weak meson-nucleon coupling
constant $f^1_\pi$. The weak meson-nucleon coupling constants have also been
calculated by Kaiser and Meissner (KM) [6] within the framework of a non-linear
chiral effective Lagrangian which includes $\pi$, $\rho$, and $\omega$ mesons. In this
model $f^1_\pi$ is considerably smaller than the ``best guess value" of DDH or FCDH.
Furthermore, a non-zero and non-negligible value for the seventh weak meson-nucleon 
coupling constant $h^{'1}_\rho$ was found. A complete determination of the
six weak meson-nucleon coupling constants requires at least six pieces of
independent experimental information. As of to date there do not exist enough
experimental constraints of statistical significance to determine the six weak
meson-nucleon coupling constants. Consequently, one needs several new precision
parity violation measurements.

   Impressively precise measurements of the longitudinal analyzing power $A_z$
in $p$--$p$ scattering have now been made at 13.6 MeV 
 [$A_z$ = (-0.93 $\pm$ 0.20 $\pm$ 0.05) $\times$ 10$^{-7}$] at the University of Bonn [7] and
at 45 MeV [$A_z$ = (-1.57 $\pm$ 0.23) $\times$ 10$^{-7}$] at the Paul Scherrer Institute
(PSI) [8]. Here $A_z$ is defined as $A_z$ = ($\sigma^+ - \sigma^-)/(\sigma^+ + \sigma^-$),
where $\sigma^+$ and $\sigma^-$ represent the scattering cross section for polarized
incident protons of positive and negative helicity, respectively, integrated
over a range of angles determined by the acceptance of the particular
experimental apparatus. A non-zero value of $A_z$ implies parity violation due to
the non-zero pseudo-scalar observable $\vec{\sigma} \cdot \vec{p}$ with $\vec{\sigma}$ 
the spin and
$\vec{p}$ the momentum of the incident proton. From the PSI measurement at 45 MeV and
the $\sqrt{E}$ energy dependence of $A_z$ at lower energies, one can extrapolate
$A_z$ at 13.6 MeV to be $A_z$ = (-0.86 $\pm$ 0.13) $\times$ 10$^{-7}$. There exists thus
excellent agreement between the above two lower energy measurements. Both
results allow pinning down a combination of the effective $\rho$ and $\omega$ weak
meson-nucleon coupling constants $h^{pp}_\rho$ and $h^{pp}_\omega$, with
$h^{pp}_\rho = h^0_\rho + h^1_\rho + \frac{h^2_\rho}{\sqrt{6}}$ and $h^{pp}_\omega$ = 
$h^0_\omega$ + $h^1_\omega$.
It should be noted that a measurement of $A_z$ in $p$--$p$ scattering is sensitive only
to the short range part of the parity violating interaction (parity violating
$\pi^0$ exchange would simultaneously imply CP violation and is therefore
suppressed). Following the approach of Adelberger and Haxton [9], one can fit
the more significant nuclear parity violation data using theoretical
constraints by the two parameters $f^1_\pi$ and ($h^0_\rho$ + 0.6 $\times$ 
$h^0_\omega$). This
leaves the experimental value of $f^1_\pi$ = $\left(0.28 {{+0.89}\atop{-0.28}}\right)$ $
\times 10^{-7}$ [10]
at the border of the deduced range so determined [11]. See Table 1, last two
columns. 
  
   A partial wave decomposition allows the various contributions to $A_z$ to be
calculated based upon reasonable estimates of the parity violating mixing
angles [12]. These mixing angles are directly related to the parity violating
transition amplitudes ($^1S_0$ - $^3P_0$), ($^3P_2$ - $^1D_2$), ($^1D_2$ - 
$^3F_2$), ($^3F_4$ - $^1G_4$), etc.
The energy dependence of the first two parity violating transition amplitudes
contributing to $A_z$ is shown in Fig.~1 [13]. For energies below 100 MeV
essentially only the first parity violating transition amplitude ($^1S_0$ - $^3P_0$)
contributes. One notices the increase in importance
of the second parity violating transition amplitude for energies
above 100 MeV. There exists a further $p$--$p$ parity violation measurement at 800
MeV with $A_z = (2.4 \pm 1.1) \times 10^{-7}$ [14]. Interpretation of the latter
result in terms of the effective $\rho$ and $\omega$ weak meson-nucleon coupling
constants is more difficult due to the presence of a large inelasticity (pion
production).

   As shown in Fig.~1 there is a unique feature at an energy of 230 MeV: the
($^1S_0$ - $^3P_0$) transition amplitude contribution integrates to zero. This reflects
a change in sign of both the $^1S_0$ and $^3P_0$ strong interaction phases near 230 MeV
and is completely independent of the weak meson-nucleon coupling constants.
The absolute scale and sign of the ordinate are determined by the weak
interaction. Neglecting a small contribution ($\sim$~5\%) from the ($^1D_2$ - $^3F_2$)
transition amplitude, a measurement of $A_z$ at 230 MeV constitutes a measurement
of the ($^3P_2$ - $^1D_2$) transition amplitude. Simonius [15] has shown that the
($^3P_2$ - $^1D_2$) transition amplitude depends only weakly on $\omega$-exchange (to an
extent dependent on the choice of the vector meson-nucleon coupling constants
of various $N-N$ potential models), whereas $\rho$-exchange and $\omega$-exchange 
contribute to the ($^1S_0$ - $^3P_0$) transition amplitude with equal weight.
Therefore, a measurement of $A_z$ at an energy of 230 MeV constitutes a
determination of $h^{pp}_\rho$.

   Various theoretical predictions of the longitudinal analyzing power have
been reported; at 230 MeV the values of $A_z$ are +0.7 $\times$ 10$^{-7}$ [16],
+0.6 $\times$ 10$^{-7}$ [13], and +0.4 $\times$ 10$^{-7}$ [17]. Extensions to the one-boson
exchange model have been made to include 2 $\pi$ and $\pi$ - $\rho$ exchanges via
$N$--$\Delta$ and $\Delta$--$\Delta$ intermediate states to which the ($^3P_2$ - $^1D_2$)
transition amplitude is particularly sensitive. For instance, Iqbal and
Niskanen [16] find that the $\Delta$ isobar contribution at 230 MeV (dependent on
$f^1_\pi$) may be as large as the single $\rho$-exchange contribution, enhancing the
value of $A_z$ by a factor of two. What is required is a self-consistent
theoretical calculation of the longitudinal analyzing power $A_z$, avoiding
possible double counting and taking into account that the value of $f^1_\pi$ is
constrained by experiment to be rather small. Considering all the above,
a measurement of $A_z$ at 230 MeV to an accuracy of $\pm 2 \times 10^{-8}$ would
provide a most important determination of parity violation in $p$--$p$ scattering.  

   In the current TRIUMF experiment, a 200 nA proton beam with a polarization
of 0.80 to 0.82, extracted from the optically pumped polarized ion source
(OPPIS), after passing\hfilneg\ \par
 
\renewcommand{\arraystretch}{2}
\begin{sidewaystable}
Table 1. \\
Weak meson-nucleon couplings constants\\
\noindent
\begin{tabular*}{20cm}{@{\extracolsep{\fill}}cr@{}c@{}lccr@{}c@{}lccccr@{}c@{}l}
\hline \hline
Coupling & \multicolumn{10}{c}{Theoretical} & \multicolumn{5}{c}{Experimental}
\\ 
\cline{1-1}\cline{2-12} \cline{13-16}
     & \multicolumn{3}{c}{range}  & `best value' & value & \multicolumn{3}{c}{range}  & `best value' & value     
     & value      & best fit & \multicolumn{3}{c}{range} \\
       & \multicolumn{3}{c}{{\sc (ddh)}} & {\sc (ddh)}  & {\sc (dz)} & \multicolumn{3}{c}{{\sc (fcdh)}} & {\sc (fcdh)}
        & {\sc (d)} & {\sc (km)} &         & \\
\hline
$f^1_\pi$    & 0     & $\to$ & 11.4 & 4.6   &  1.1  & 0     & $\to$ & 6.5  &  2.7 &  2.7 &  0.19 &  2.3& 0 &$\to$& 11  \\
$h^0_\rho$   & -31   & $\to$ & 11.4 & -11.4 & -8.4 & -31   & $\to$ & 11   & -3.8 & -6.1 & -1.9  & -5.7 &-31 &$\to$& 11  \\
$h^1_\rho$   & -0.38 & $\to$ & 0    & -0.19 &  0.38 & -1.1  & $\to$ & 0.4  & -0.4 & -0.4 & -0.02 & -0.2&-0.4 &$\to$& 0.0 \\
$h^2_\rho$   & -11.0 & $\to$ & -7.6 & -9.5  & -6.8 & -9.5  & $\to$ & -6.1 & -6.8 & -6.8 & -3.8  & -7.6 &-11 &$\to$&-7.6 \\
$h^0_\omega$ & -10.3 & $\to$ &  5.7  & -1.9 & -3.8 & -10.6 & $\to$ & 2.7  & -4.9 & -6.5 & -1.1  & -4.9 &-10 &$\to$& 5.7 \\
$h^1_\omega$ & -1.9  & $\to$ & -0.8 & -1.1  & -2.2 & -3.8  & $\to$ & -1.1 & -2.3 & -2.3 & -1.0  & -0.6 &-1.9 &$\to$& -0.8 \\
\hline \hline   
\end{tabular*}
\end{sidewaystable}

\begin{center}
\epsfig{figure=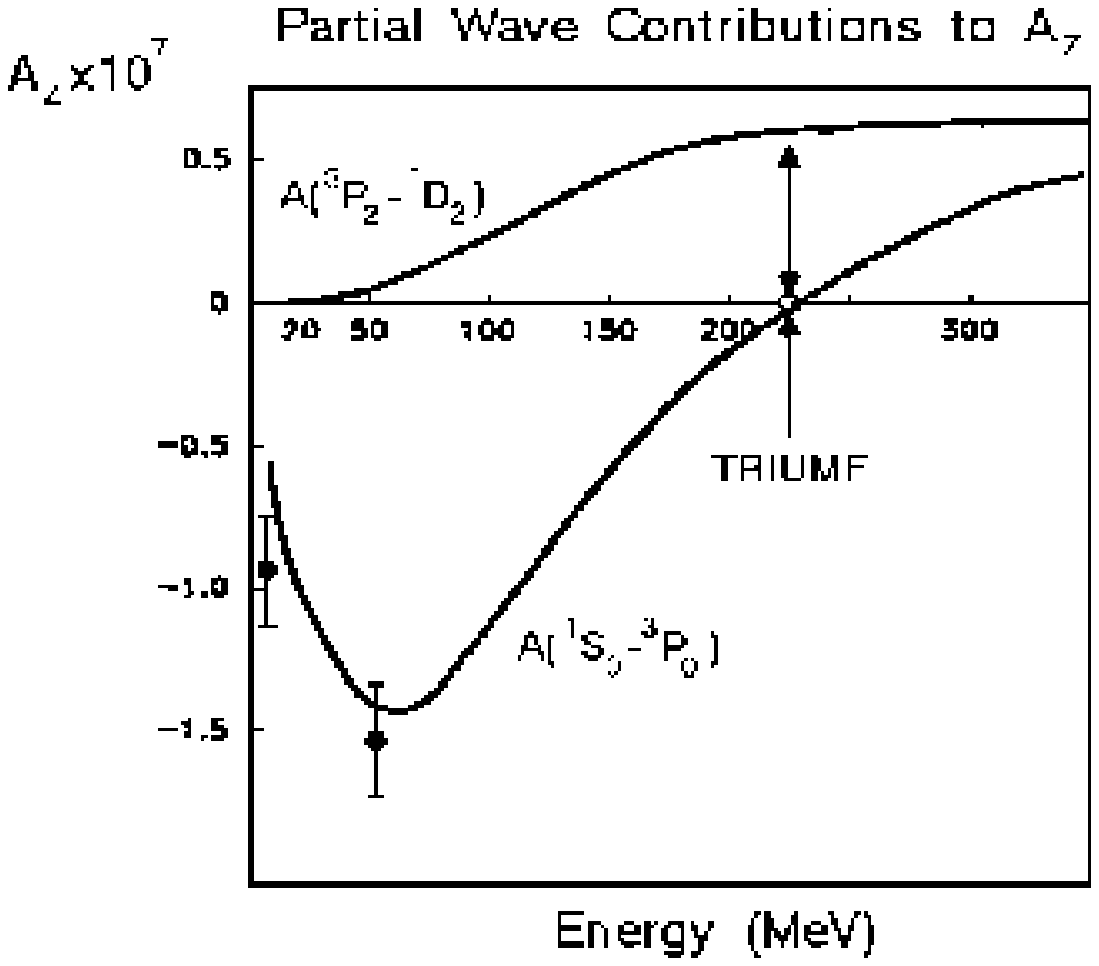,height=7.8cm}
\end{center}
\noindent
\parbox{16cm}{Figure 1. Partial wave decomposition of the $p$--$p$ parity violating longitudinal
    analyzing power $A_z$ [Ref.13].}
\vspace{0.2cm}

\noindent
a Wien filter in the injection line is accelerated
through the cyclotron to an energy of 221.3 MeV. A combination of
solenoid-dipole-solenoid-dipole magnets on the external beam line provides a
longitudinally polarized beam with positive or negative helicity. The
longitudinal analyzing power $A_z$ follows from the helicity dependence of the $p$--$p$
total cross section as determined in precise measurements of the normalized
transmission asymmetry through a 0.40 m long liquid hydrogen (LH$_2$) target:
$A_z$ = $-(1/P)(T/S)(T^+ - T^-)/(T^+ + T^-)$, where $P$ is the incident beam longitudinal
polarization, $T = 1 - S$ is the average transmission through the target, and the
+ and -- signs indicate the helicity state.

   There are many other effects that can cause a helicity correlated change in
transmission. Very strict constraints are imposed on the incident polarized
beam in terms of helicity correlated changes in intensity, in transverse x
(horizontal) and y (vertical) beam position and direction, in beam width
(given by $\sigma_x$ and $\sigma_y$), and in energy, and further transverse
polarizations (given by $P_x$ and $P_y$), and first moments of
transverse polarizations (given by $<xP_y>$ and $<yP_x>)$, together with deviations
of the transmission measuring apparatus from cylindrical symmetry. Systematic
errors arising from the imperfections of the incident beam and the response of 
the transmission measuring apparatus are individually not to exceed one tenth of
the expected value of $A_z$ (or 6 $\times$ 10$^{-9}$). Potentially particular troublesome
are residual transverse polarizations and their first moments (so called
``circulating" polarization profiles), as well as helicity correlated energy
changes. Uncorrelated fluctuations contribute to the rms noise in the
measurement and increase the total data taking time. In addition to imposing
strict constraints on the incident beam and on the measurement apparatus in
order to reduce systematic errors, the approach which is being followed is to
further measure the sensitivity or response to residual imperfections, to
monitor these imperfections during data taking and to make corrections as
appropriate. 

   Helicity changes are implemented through shifts in the linearly polarized
laser light frequency, minimizing helicity correlated changes in the
accelerated beam properties. The beam parameters are selected to produce an
achromatic waist at the LH$_2$ target, circular in cross section 
($\sigma_x$~=~$\sigma_y$~=~6~mm). To negate possible helicity correlated changes in the incident
beam energy, the currents in the two solenoid magnets as well as the rotation
angles around the beam axis of three sets of quadrupole magnets 
 are switched in polarity once every three days. This allows
for a linear combination of all four permutations of helicity states from 
the polarized ion source to the parity violation measurement apparatus in the
deduction of $A_z$. In the cyclotron the spin direction is either parallel or
antiparallel to the guide magnetic field. Helicity changes are made in a
semi-random eight-state cycle, which is designed to cancel out slow drifts in
various beam properties.

   Producing a longitudinally polarized beam for the parity violation
experiment requires careful tuning of the polarized ion source, of the injection
beam line, of the acceleration through the cyclotron, and of the beam transport
line to the parity violation measuring apparatus. Many improvements to the
operation of the polarized source had to be made to reduce helicity correlated
changes in intensity, emittance, polarization, and energy. The Faraday effect
provides a means to monitor and control the polarization of the Rubidium vapour
(source of the polarized electrons which are exchanged with the passing
protons) on line using a probe laser. The polarization direction of the linearly
polarized probe laser light is rotated by an angle proportional to the 
Rubidium vapour polarization. The polarizations of the Rubidium vapour in the
two helicity states are maintained to be the same within 0.5\% close to 100\% .
The Faraday rotation measurement also provides confirmation of the helicity
state at the polarized ion source.

   To aid in tuning, various retractable horizontal and vertical wire chambers
were placed along the beam line. Following the second solenoid magnet, where
the polarization direction has both a longitudinal and a horizontal sideways
component, a four branch polarimeter measures the transverse polarization
components, while a beam energy monitor measures the relative energy of the
proton beam with a precision of $\pm$20 keV during a one hour data taking run.
The absolute energy has to agree within a few MeV with the energy for which the
contribution of the ($^1S_0$ - $^3P_0$) transition amplitude integrates to zero, taking
into account the finite geometry of the transmission measuring apparatus; but
any changes in energy greater than 40 keV will introduce transverse
polarization components in excess of 0.001 for the canonical setting of all
beam transport magnetic elements. All beam transport elements have their
excitations monitored on a continuous basis (superconducting solenoids -
currents; dipole magnets using NMR probes; quadrupole magnets using Hall
probes).

   Figure 2 provides a diagram of the downstream part of the experimental
setup. The longitudinally polarized beam, incident from the lower right, passes
first a series of diagnostic devices - a set of three beam intensity profile
monitors (IPMs), and a pair of transverse polarization profile monitors (PPMs) -
 before reaching the LH$_2$ target which is preceded and followed by transverse
electric field ionization chambers (TRICs) to measure the beam current. 

   The IPMs, which are based on secondary electron emission from thin, 3~$\mu$m
thick nickel foils placed between 8~$\mu$m aluminum high voltage foils, measure
the beam intensity profile\hfilneg\ \par
\newpage

\begin{center}
\begin{turn}{90}
\epsfig{figure=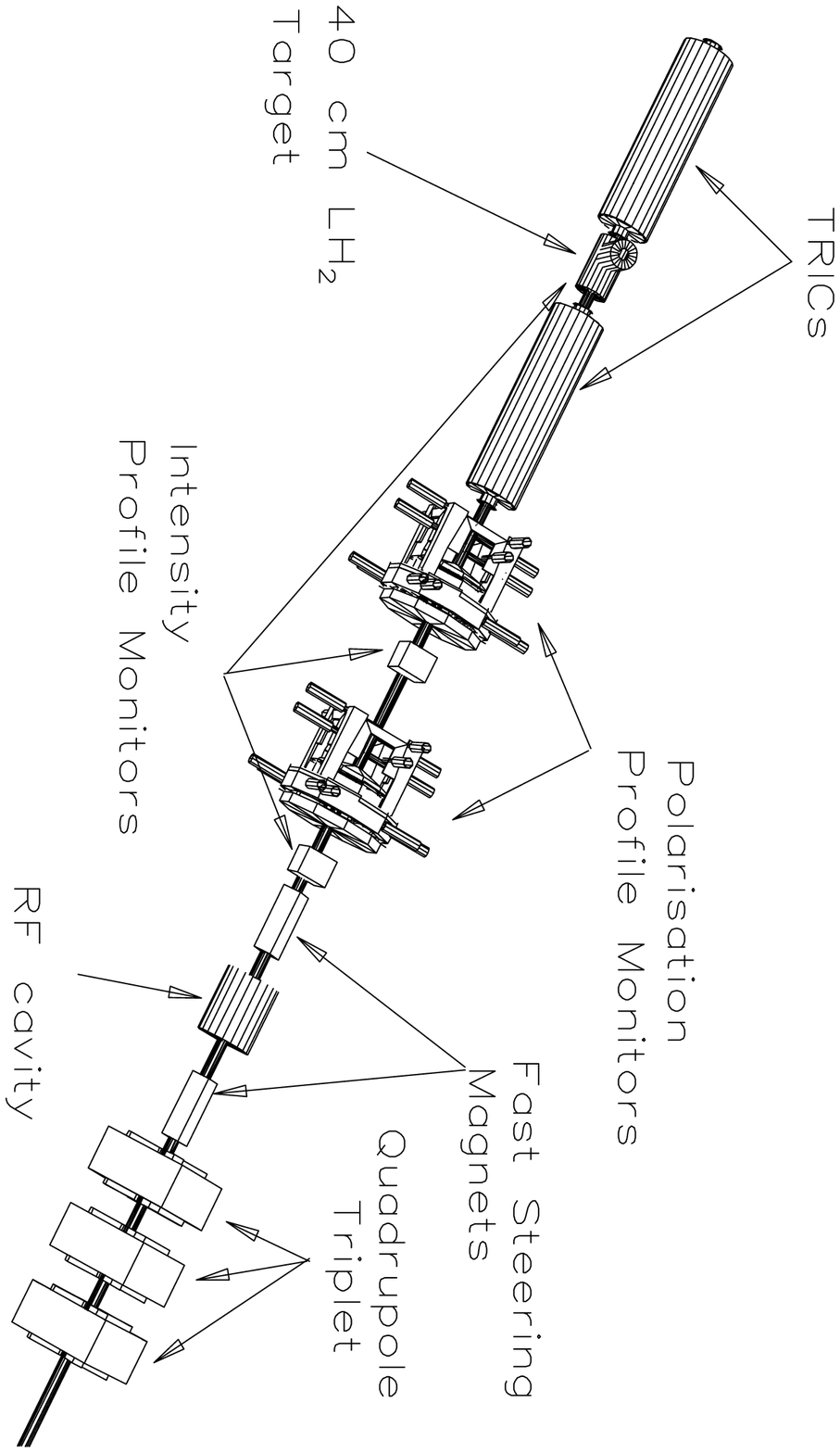,height=10cm} 
\end{turn}
\end{center}

\noindent
Figure 2. Three dimensional view of the TRIUMF parity violation detection apparatus.
\vspace{0.2cm}

\noindent
with harps of 31 foil strips (1.5 mm wide, separated
2.0 mm center to center) in both the
vertical (x-profile) and horizontal
(y-profile) directions. The third IPM placed just in front of the LH$_2$ target
has 10~$\mu$m thick nickel strips (2.5 mm wide, separated by 3.0 mm center to
center). Beam centroid evaluators determine the beam intensity profile
centroids on line at two locations through appropriate integration of the
discrete distributions; a corresponding normalized error signal is used to drive
feedback loops to x and y, ferrite-cored fast steering magnets. This allows the beam intensity
profile centroids to be kept fixed within 1~$\mu$m with an offset less than
20~$\mu$m from the ``neutral axis" in both x and y during a one hour data taking
run. Typical beam intensity profile widths are: IPM-1 $\sigma_x$ = $\sigma_y$ 
= 5 mm; IPM-2 : 4 mm; IPM-3 : 6 mm. Sensitivities to helicity correlated position and
size modulations are determined with enhanced modulations synchronized to the
helicity sequence for the experiment. Position and size modulations are
measured during the course of data taking to allow for offline corrections.
Typical values of these modulations in a one hour data taking run are
$\Delta x$, $\Delta y$ $<$ (0.5 $\pm$ 0.3) $\mu$m and 
$\Delta \sigma_x$, $\Delta\sigma_y$ $<$ (1.0 $\pm$ 0.6) $\mu$m.
 
   The PPMs are based on $p$--$p$ scattering using CH$_2$ targets. Scattered protons
are detected in a forward arm at 17.5$^\circ$ with respect to the incident beam
direction with a pair of scintillation counters. The solid angle defining
scintillator is rotated around an axis perpendicular to the scattering plane to
compensate for changes in solid angle and differential cross section when the
CH$_2$ target blade moves through the beam (see Fig.~3). Coincident recoil protons
are detected in a backward arm at 70.6$^\circ$ with respect to the incident
beam direction with a recoil scintillation counter; a second scintillation
counter acts as a veto for higher energy protons from $^{12}$C($p,p$)$X$. Each PPM
contains detector assemblies for  ``left" scattered protons, ``right" scattered
protons; ``down" scattered protons, and ``up"
scattered protons. The targets
consist of CH$_2$ blades, 1.6 mm wide and 5.0 mm deep along the incident beam
direction. The blades move through the beam on a circle of 0.215 m at a 
frequency of 5 Hz. Each PPM has four blades: two which scan the polarization
profile in the horizontal direction and allow for determining the quantity
(L - R)/(L + R) and therefore $P_y$ as function of x for each of the two helicity
states, and two which scan the polarization profile in the vertical direction
and allow for determining the quantity (D - U)/(D + U) and therefore $P_x$ as
function of y for each of the two helicity states. Residual\hfilneg\ \par
\newpage

\begin{center}
\epsfig{figure=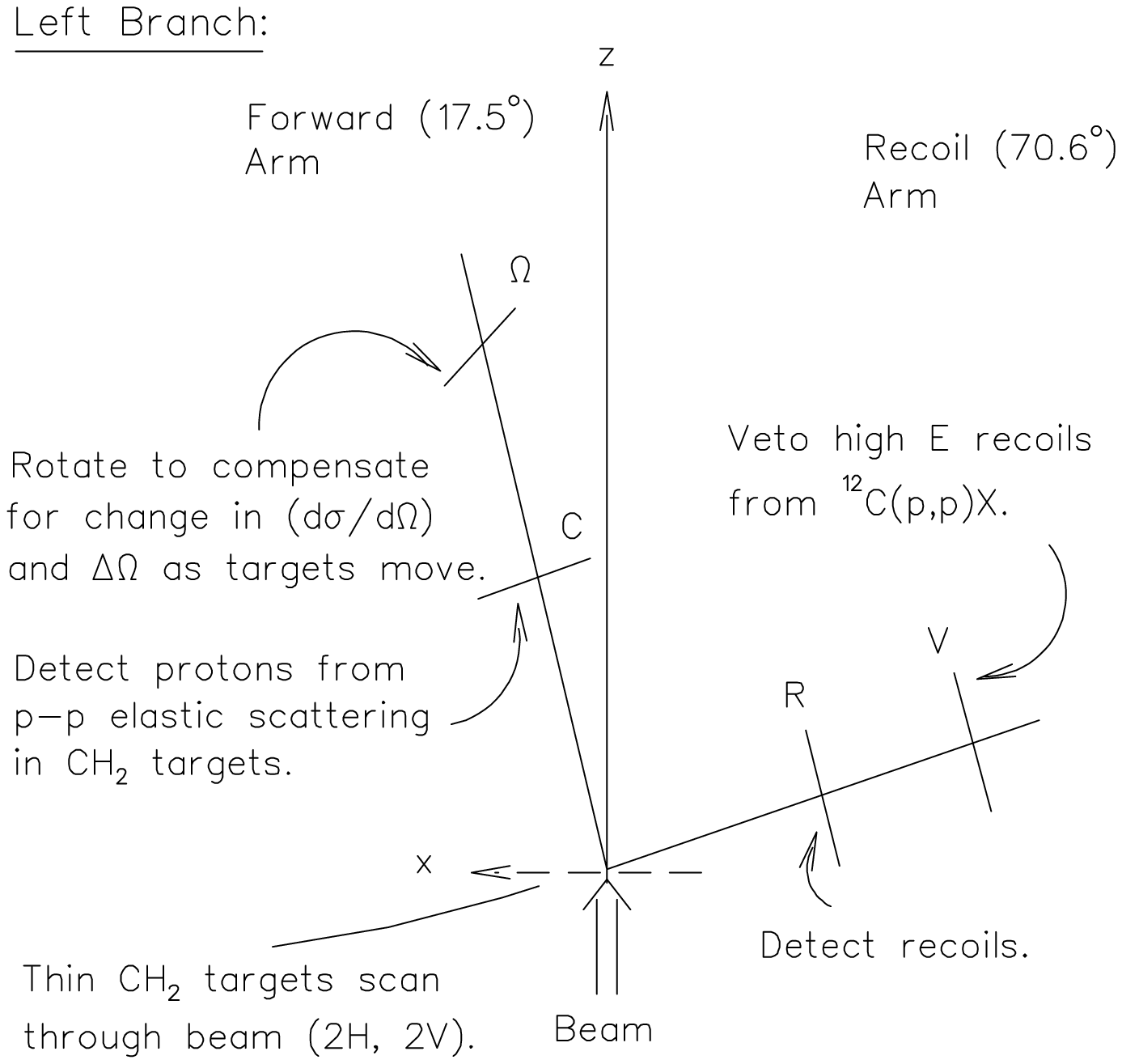,height=9.73cm}
\end{center}

\noindent
Figure 3. Schematic representation of one of the four detector assemblies of each PPM.
\vspace{0.2cm}

\noindent
transverse
polarizations (which change sign with helicity reversals) can cause a false $A_z$
via the parity allowed transverse analyzing power which produces asymmetric
scattering in the LH$_2$ target. The sensitivities to residual transverse
polarizations are dependent on the incident beam position and on the geometry 
of the parity violation detection apparatus. Both the sensitivities and the
``neutral axis" are determined by introducing enhanced transverse polarizations
$P_x$ and $P_y$. The first moments of the transverse polarizations, $<xP_y>$ and 
$<yP_x>$,
can arise from an inhomogeneous polarization distribution of the cyclotron
beam at the stripper foil location and spin precession in the magnetic field
gradients at the entrance and exit of solenoids, dipole magnets, and quadrupole
magnets. The first moments of the transverse polarizations are 
$\leq (20 \pm 5)$~$\mu$m as measured in a one hour data taking run. The first moments scale with
beam size and are minimal at a waist of the incident proton beam. Consequently,
the beam transport parameters were chosen so as to have a waist of the incident
proton beam at the LH$_2$ target. 

   With two PPMs each with four blades, the spin flip or helicity change
becomes 40 Hz, i.e., in one cycle all eight blades of the two synchronized PPMs
(four blades of PPM-1 and then four blades of PPM-2) will have passed once
through the beam. The master clock for sequencing the whole experiment,
including helicity changes, is derived from the PPM shaft encoders. In order
to suppress the effects of uncorrelated fluctuations in the beam properties
up to second order, a cycle consists of the following sequence of helicities:
$+ - - + - + + -$, lasting 200 ms. Eight cycles are repeated, with the first
helicity chosen so as to form an eight by eight symmetric matrix of helicity
states (designated a super-cycle). After four such super-cycles, the pumping
laser light is blocked by a shutter during one super-cycle for control
measurements. In every other super-cycle of the latter, the sensitivity to
helicity correlated changes in the incident beam intensity is measured using
a fourth laser which through photodetachment removes up to 0.2\% of the H$^-$ ion
beam in the injection line with the 40 Hz frequency. Helicity correlated
changes in the incident beam intensity cause a false parity violation signal
due to nonlinearities of the TRICs and the associated electronics. Precision
analog subtraction of the two TRIC currents minimizes the sensitivity of $A_z$ to
$\Delta I/I$. Careful tuning of OPPIS gives values of
$\Delta I/I$ = (2 $\pm$ 1) $\times$ 10$^{-5}$ typically in a one hour data taking run.
Coherent intensity modulations of up to 0.2\% allow for tuning of the analog
subtractor for minimum sensitivity at 200 nA and for determining the sensitivity
for offline correction of the data. Of each helicity state
of 25 ms duration a little more than 1 ms is reserved for the polarization
to stabilize following a helicity change, the next 6.4 ms is reserved for the
measurement of the transverse polarization (one of the CH$_2$ blades whisking
through the beam), and the following 1/60 sec is used for the actual parity
violation measurement (determining the helicity dependent transmission).
A small phase slip is introduced so that the master clock and the line frequency
are again precisely in phase after 20 minutes.

   The LH$_2$ target has a flask of 0.10 m diameter and a length of 0.40 m.
Special precautions have been taken to make the end windows of the target flask
perfectly flat and parallel. Maximum heat load of the target is 25 W with
operation at 5 W. By circulating the liquid hydrogen rapidly, density gradients
are minimized. The target flask is movable remotely within $\pm$~5 mm in two
orthogonal directions at both ends to position it on the ``neutral axis". The
total scattering probability at 221.3 MeV by the 0.40 m long LH$_2$ target is
close to 4\%. The target flask length is limited by multiple Coulomb scattering
considerations; the various entrance and exit windows (and all energy
degrading foils in the beam) are kept to the minimally allowable thickness.

   The main detectors are two transverse electric field ionization chambers,
producing current signals due to direct ionization of the ultra-high purity
hydrogen gas by the beam. Field shaping electrodes plus guard rings ensure a
0.15 m wide, by 0.15 m high, by 0.60 m long sense region between the parallel
electrodes (negatively charged cathode and signal plate), with the field lines
all parallel and perpendicular to the electrodes. The TRICs have been designed
for operation at -35 kV at one atmosphere; in practise they are operated at a
pressure of 150 torr and a high voltage of -8 kV. The entrance and exit
windows are located at approximately 0.9 m from the center of the TRICs to
range out spallation products from proton interactions with the stainless steel
windows. The design of the TRICs incorporated considerations of noise due to
$\delta$-ray production and due to recombination.  

   The proton beam energy in the downstream TRIC is on average 27 MeV lower
than in the upstream TRIC due to energy loss in the LH$_2$ target. Helicity
correlated energy modulations will cause a false $A_z$ due to the energy
dependence of the energy loss in the hydrogen gas of the TRICs. The sensitivity
to coherent energy modulations was determined using a RF accelerating cavity
placed upstream of IPM-1 in the beam line. The RF cavity could produce coherent
energy modulations up to 1000 eV in the 221.3 MeV proton beam. The measured
sensitivity of $(2.9 \pm 0.3) \times 10^{-8}$ eV$^{-1}$ agrees very well with the
prediction obtained in Monte Carlo simulations of the experiment of
$2.8 \times 10^{-8}$ eV$^{-1}$ and places a very stringent constraint on the
maximally allowable helicity correlated energy modulation of the incident proton
beam. Since such an energy modulation cannot be measured directly during
parity violation data taking, the false $A_z$ needs to be removed through an
appropriate linear combination of the data taken with all possible helicity
combinations and spin precessions. Helicity correlated energy modulations
can originate in OPPIS and during acceleration in the cyclotron. Amplification
of the former during acceleration in the cyclotron is reduced significantly
using a position feedback system in the injection beam line. The naturally
occurring coherent energy modulation in OPPIS is measured regularly and is
typically less than 20 meV.

   The achieved quality of the polarized proton beam at the parity violation
data taking apparatus and the performance of the latter will allow for a
statistical precision in $A_z$ of $\pm 2 \times 10^{-8}$ in about 300 hours. One fifth
of the required data was acquired during four weeks in February-March 1997.
A preliminary result for the longitudinal analyzing power at 221.3 MeV is 
$A_z$ =
(1.07 $\pm$ 0.41 $\pm$ 0.37) $\times$ 10$^{-7}$, where the first error is statistical and
the second error represents the systematic uncertainty. The result is shown
together with the theoretical prediction of Driscoll and Miller [17] in Fig.~4.
Even though the
theoretical prediction overestimates the size of $A_z$ at the
lower energies, it shows the
expected energy behavior. Data taking for the 
221.3 MeV experiment will continue during several scheduled running periods.

   A further $p$--$p$ parity violation experiment is being prepared at TRIUMF at an
energy of 450 MeV [18]. This measurement can be made with minimal changes to
the apparatus used in the present TRIUMF $p$--$p$ parity violation experiment.
It is intended to arrive at the same overall uncertainty of $\pm 2 \times 10^{-8}$.
The combination of measurements at 221 Mev and 450 MeV would give an
independent determination of the weak meson-nucleon coupling constants $h^{pp}_\rho$
and $h^{pp}_\omega$. Measurements of $A_z$ in $p$--$p$ scattering are also planned at COSY of
the Kernforschungsanlage J\"ulich near 230 MeV and with an extension to 1.3 GeV
[19]. The choice of the latter energy (or better a higher energy) is in part
motivated by the earlier 5.13 GeV measurement of the longitudinal analyzing
power $A_z$ (on a water target) at the ZGS of Argonne National Laboratory, which
resulted in $A_z = (26.5 \pm 6.0 \pm 3.6) \times 10^{-7}$ [20]. The result is an order of
magnitude larger than what is expected using conventional scaling arguments.
It must be remarked that various reevaluations of the experiment have not
discovered any flaw, in the way the experiment was conducted, that could have
led to such a false large $A_z$. It has also been pointed out that when Glauber
shadowing is taken into account, the fundamental $p$--$p$ parity violating $A_z$
increases by as much as 40\% [21]. Figure 5 shows the energy dependence of $A_z$;
note the logarithmic scale used for the abscissa. The theoretical prediction
for the higher energies [23] was normalized to the 5.13 GeV datum. This
calculation is based on a di-quark model introducing a parity violating
component of the nucleon wave function. Goldman and Preston find an important role for
diagrams in which the weak interaction between the members of a vector di-quark
in the polarized proton is accompanied by the strong interaction between that
di-quark and a quark of the other proton. The theoretical prediction agrees
with the 800 MeV experimental datum and is characterized by a steep increase 
with increasing energy. It predicts a value for $A_z$ at 20 GeV of the order
10$^{-5}$. Note that a 200 GeV experiment has placed an upper limit (95\% C.L.)
on the $p$--$p$ longitudinal analyzing power $A_z$ of $5.7 \times 10^{-5}$ [24]. The
theoretical interpretation of the unexpectedly large result at 5.13 GeV has
created a great deal of controversy [25]. Clearly, the 5.13 GeV result presents
a great challenge both in obtaining experimental confirmation through a new\hfilneg\ \par


\begin{center}
\epsfig{figure=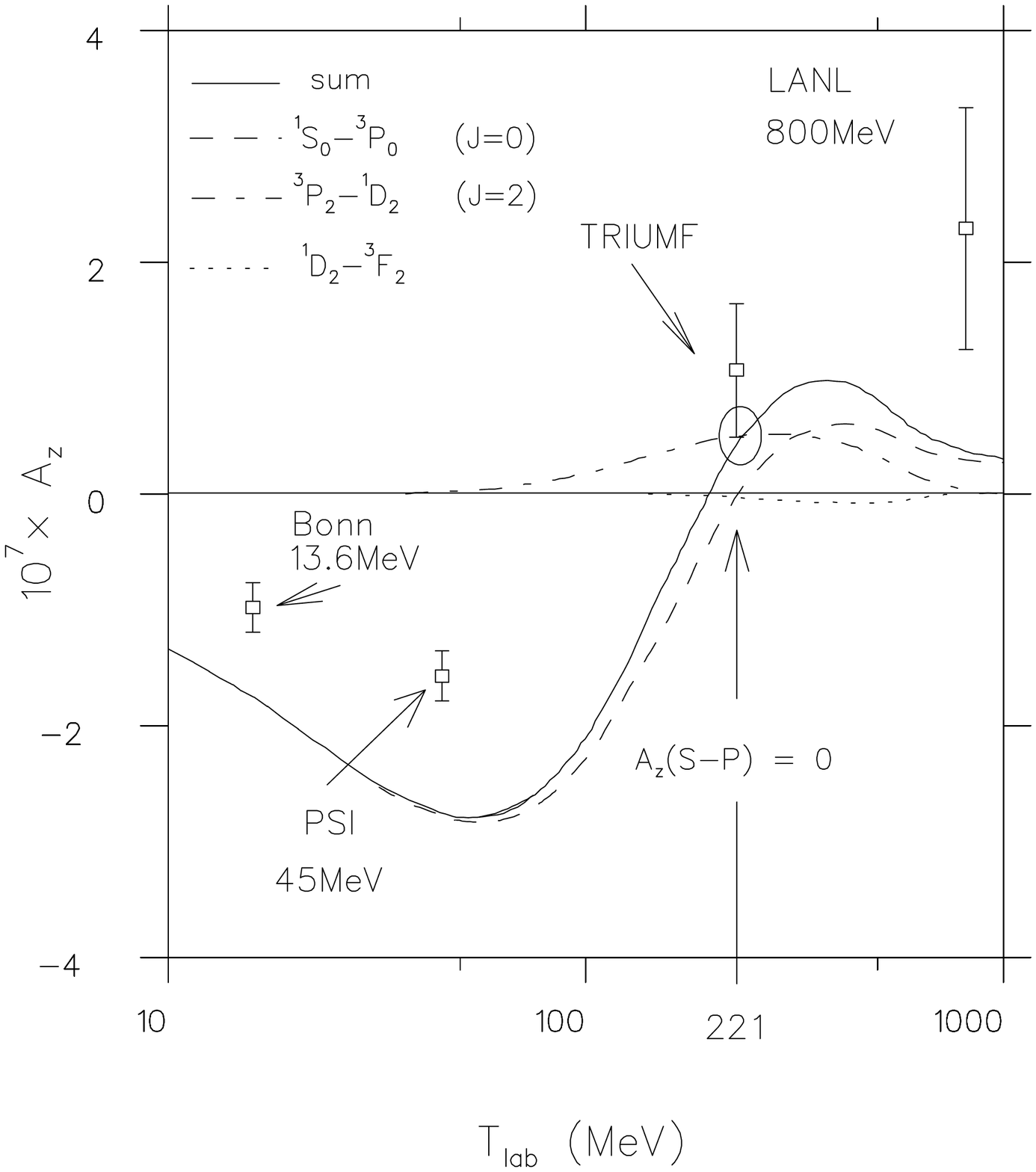,height=9cm}
\end{center}
\noindent
\parbox{16cm}{
Figure 4. Partial wave decomposition giving the first three contributions to $A_z$ as
    given by Driscoll and Miller [Ref.17] compared to the experimental data. Note the zero-crossing of the
    ($^1S_0$ - $^3P_0$) transition amplitude contribution.}


\begin{center}
\epsfig{figure=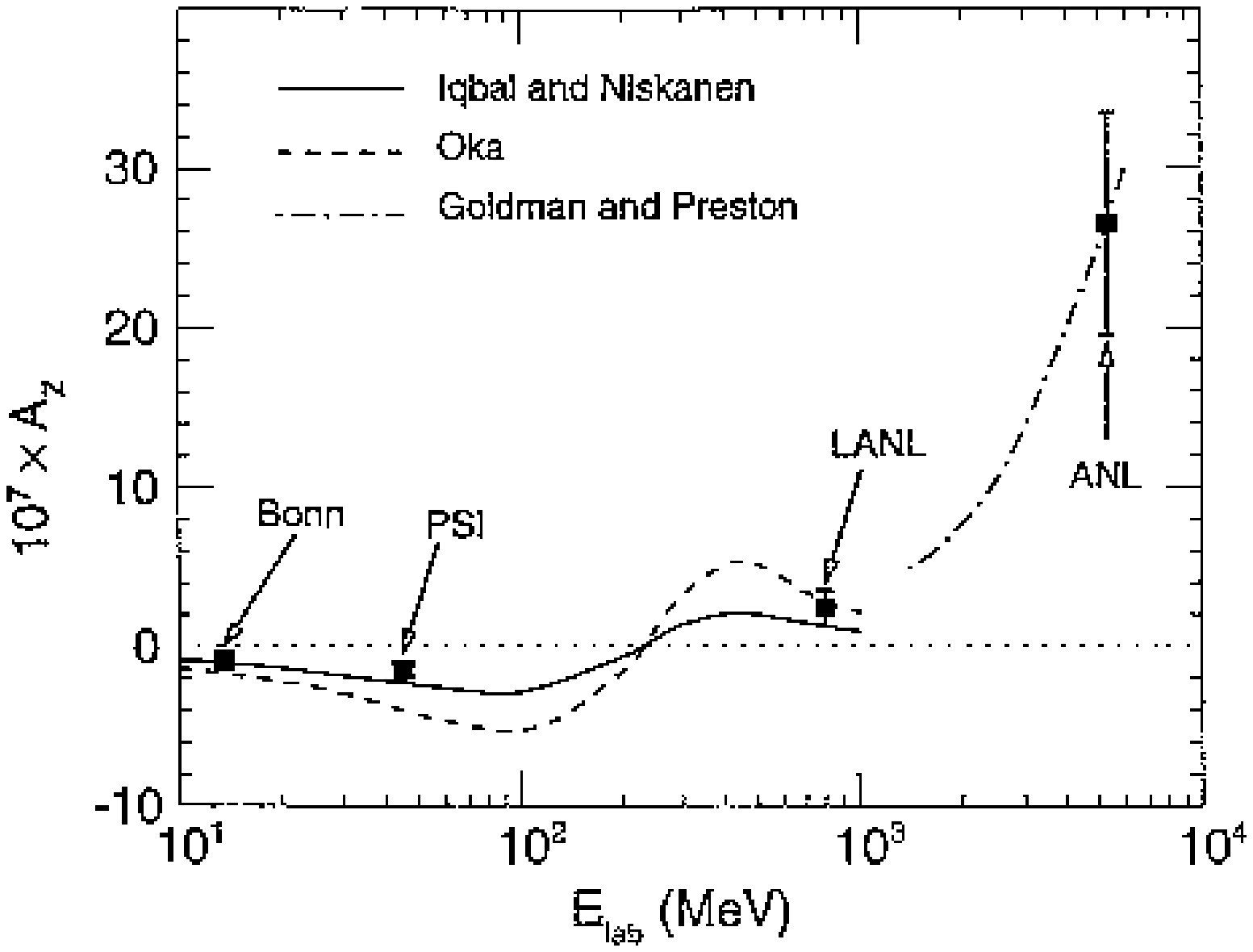,height=7.8cm}
\end{center}
\noindent
\parbox{16cm}{
Figure 5. Energy dependence of the $p$--$p$ parity violating longitudinal analyzing power 
    $A_z$. The solid and dashed curves represent theoretical models
    [Ref. 16 and 22] based on weak meson exchange; the dot-dashed curve
    is described in the text.}
    \vspace{0.2cm}


\noindent
measurement at 5 GeV and in obtaining a self-consistent theoretical
explanation. If confirmed experimentally, there is a need for a further
experiment at an energy of tens of GeV either in a fixed target arrangement or
in a storage ring environment.

\vspace{0.2cm}
\bibliographystyle{unsrt}

\begin{thebibliography}{99}
\bibitem[*]  \protect Work supported in part by the Natural Sciences and Engineering Research
 Council of Canada
\bibitem{1} B.~Desplanques, J.F.~Donoghue,and B.R.~Holstein, Ann. Phys. (N.Y.) 124, 449 (1980).
\bibitem{2} V.M.~Dubovik and S.V.~Zenkin, Ann. Phys. (N.Y.) 172, 100 (1986).
\bibitem{3} G.B.~Feldman, {\em et al.}, Phys. Rev. C43, 449 (1980).
\bibitem{4} B.~Desplanques, Nucl. Phys. A335, 147 (1980).
\bibitem{5} B.~Desplanques, in Proceedings of the International Workshop on Parity
    Violation and Time Reversal Invariance, ed. N.~Auerbach and J.D.~Bowman (World Scientific,
    Singapore, 1996), p.~98           
\bibitem{6} N.~Kaiser and U.G.~Meissner, Nucl. Phys. A499, 699 (1989); A510, 759
(1990).
\bibitem{7} P.D.~Eversheim, {\em et al.}, Phys. Lett. B256, 11 (1991); P.D.~Eversheim, 
private communication (1994).
\bibitem{8} S.~Kistryn, {\em et al.}, Phys. Rev. Lett. 58, 1616 (1987).
\bibitem{9} E.G.~Adelberger and W.C.~Haxton, Ann. Rev. Nucl. Part. Sci. 35, 501 (1985).
\bibitem{10}W.~Haeberli and B.R.~Holstein, in Symmetries and Fundamental Interactions in
    Nuclei, ed. W.C.~Haxton and E.M.~Henley (World Scientific, Singapore, 1995), p.~17.
\bibitem{11} H.C.~Evans, {\em et al.}, Phys. Rev. Lett. 55, 791 (1985); S.A.~Page, {\em et al.}, 
Phys. Rev. C35, 1119 (1987); M.~Bini, {\em et al.}, Phys. Rev. Lett. 55, 795 (1985);
Phys. Rev. C 38, 1195 (1988).
\bibitem{12} J.~Birchall, G.~Roy, and W.T.H.~van Oers, Phys. Rev. D37, 1769 (1988).
\bibitem{13} M.~Simonius, in Proceedings of the Symposium/Workshop on Spin and Symmetries, ed. 
             W.D.~Ramsay and W.T.H.~van Oers, Can. J. Phys. 66, 245 (1988); 
             F.~Nessi-Tedaldi and M.~Simonius, Phys. Lett. B215, 159 (1988).
\bibitem{14} V.~Yuan, {\em et al.}, Phys. Rev. Lett. 57, 1680 (1986).
\bibitem{15} M.~Simonius, in Interaction Studies in Nuclei, ed. H.Jochim and B.Ziegler
    (North Holland, Amsterdam, 1975), pp.~3; in High Energy Physics with Polarized
    Beams and Targets, ed. C.~Joseph and J.~Soffer (Birkhauser Verlag, Basel, 1981), p.~355.
\bibitem{16} M.J.~Iqbal and J.~Niskanen, Phys. Rev. C42, 1872 (1990); private communication (1994).
\bibitem{17} D.E.~Driscoll and G.A.~Miller, Phys.~Rev.~C39, 1951 (1989).
\bibitem{18} TRIUMF Proposal E761, J.~Birchall, S.A.~Page, and W.T.H.~van Oers, cospokespersons.
\bibitem{19} P.D.~Eversheim, {\em et al.} in High Energy Spin Physics, ed. K-H.~Althoff and
    W.~Meyer (Springer Verlag, Berlin, 1991), p.~573.
\bibitem{20} N.~Lockyer, {\em et al.}, Phys.~Rev. D30, 860 (1984).
\bibitem{21} L.L.~Frankfurt and M.I.~Strikman, Phys. Lett. 107B, 99 (1981).
\bibitem{22} T.~Oka, Prog. Theor. Phys. 66, 977 (1981).
\bibitem{23} T.~Goldman and D.~Preston, Phys. Lett. B168, 415 (1986).
\bibitem{24} D.P.~Grosnick, {\em et al.}, Phys. Rev. D55, 1159 (1997).
\bibitem{25} M.~Simonius and L.~Unger, Phys. Lett. B198, 547 (1987); T.~Goldman, in Future
    Directions in Particle and Nuclear Physics at Multi-GeV Hadron Facilities,
    ed. D.F.~Geesaman (Brookhaven National Laboratory Report BNL-52389,1993),
    p.~140; M.~Simonius, {\em ibid.}, p.~147.     
\end{thebibliography}


\end{document}